\documentclass[12pt]{article}
\usepackage[utf8]{inputenc}
\usepackage{amsmath,amsfonts,amsmath}
\usepackage{graphicx,psfrag,epsf,epstopdf}
\usepackage{enumerate}
\usepackage{bbm}
\usepackage{natbib}
\usepackage{url,hyperref}
\usepackage{booktabs}
\usepackage[table,xcdraw]{xcolor}
\usepackage{rotating}
\usepackage[font=small]{caption}

\usepackage{algorithm}
\usepackage{algpseudocode}

\newcommand{\blind}{0}

\begin{document}

\def\spacingset#1{\renewcommand{\baselinestretch}%
{#1}\small\normalsize} \spacingset{1}

%%%%%%%%%%%%%%%%%%%%%%%%%%%%%%%%%%%%%%%%%%%%%%%%%%%%%%%%%%%%%%%%%%%%%%%%%%%%%%

\if0\blind
{
  \title{\bf A depth-based method for \\ functional time series forecasting} %with dynamic updating}
  \author{Antonio El\'ias\thanks{Supported by the Spanish Ministerio de Educaci\'on, Cultura y Deporte under grant FPU15/00625.}  \ and Ra\'ul Jim\'enez\thanks{Partially supported by the Spanish Ministerio de Econom\'ia y Competitividad under grant ECO2015-66593-P.}\hspace{.2cm} \\
Department of Statistics, Universidad Carlos III de Madrid}
  \maketitle
} \fi

\if1\blind
{
  \bigskip
  \bigskip
  \bigskip
  \begin{center}
    {\LARGE\bf Depth-based forecasting}
\end{center}
  \medskip
} \fi

\bigskip
\begin{abstract}

An approach is presented for making predictions about functional time series. %with dynamic updating.
%These are functional data obtained by slicing an almost continuous time record into natural consecutive periods. %, when the last of them has not yet been fully observed.
%The forecasting problem the last period when it has been only partially observed.
The method is applied to data coming from periodically correlated processes and electricity demand, obtaining accurate point forecasts and narrow prediction bands that cover high proportions of the forecasted functional datum, for a given confidence level.
%The trade-off between covered proportion and band width is addressed with graphical tool from which the practitioner may decide 
The method is computationally efficient and substantially different to other functional time series methods, offering a new insight for the analysis of these data structures. %, which are based on statistical models of temporal correlation.

\end{abstract}

\vfill

\newpage
\spacingset{1.45}
\section{Introduction}
\label{sec:intro}

The concept of depth for functional data has received a great deal of attention since it was introduced by \cite{Fraiman2001}.
It has been used for several applications; including classification
\citep{saraLopezJuanRomo2006, Cuevas2007, cuestaalbertos2008, Sguera2014, Hubert2017, Mosler2017},
outlier detection 
\citep{Febrero2008, IevaPaganoni2013, arribasromo2014, Chiou2014, Narisetty2015, Nagy2017},
populations comparison 
\citep{pintadoromo2009, lopezpintado2010, Nicholas2015},
and clustering \citep{kwon2015, Singh2016,Tupper2017}.
%https://dl.acm.org/citation.cfm?id=2954724
Functional versions of boxplots and other graphical tools based on different depths have been also proposed for visualizing curves with the aim of discovering features from a sample that might not be apparent by using other methods \citep{Hyndman2010, sungenton2011, Serfling2017}.
These methods are based on the so called \emph{central regions} that we combine here with some new ideas.

This paper addresses the problem of making predictions about functional time series.
These data structures come from dividing an almost continuous time record into curves corresponding to natural consecutive periods, for example days.
The forecasting problem in such as context has been an important field of research that has produced diverse seminal literature \citep{antoniadis2006, hyndman2007, aneirosperez2008, hyndman2008, aneirosperez2011}.
The prediction of a curve segment corresponding to an unobserved interval at the most recent period has been termed forecasting with dynamic updating  \citep{hyndmanshang2011}.
We use this context for presenting our approach.
Unlike previous methods, we attempt to capture the morphology of the curve to predict without using any statistical model about temporal correlation among periods.
We only suppose that the functional time series exhibits certain periodic structure. Then, past periods provide a library where we can search for some similarities with the most recent data.

The method is based on a selection of past curves that makes the segment of the most recent period a deep datum.
We show that the bands delimited by the deepest curves from the selection cover high proportions of the curve segment to predict for a given confidence level. We also provide a graphical tool for choosing how many curves to use, by taking into account the resulting band width and the proportion of time that the predicted curve will be into such band, for a given confidence level.
From the entire selected curves, we also provide point forecast.
The method is tested with simulated data and a real case study.

%This paper is organized as follows. 
%First, we review the central regions used in this paper. 
%Second, we describe the curves selection method and its use in forecasting.
%Then, we present results obtained from simulations and the Spanish electricity demand study case.
%To end, some conclusions are discussed. 

\section{Description of the Method}
\label{sec:FCR}

Let $Y$ be an almost continuous periodic time series of period $p$.
Consider the curves $y_1,y_2,\dots$ obtained by slicing $Y$ into periods, this is
\begin{equation}
y_i(t) = Y(t+(i-1)p ),  \  0\leq t \leq p, \ i =1, 2,\dots.
\end{equation}
When $y_1,...,y_n$ are observed on $[0,p]$, but $y_{n+1}$ only on $[0,q]$ (with $q<p$), and we are interested in predicting $y_{n+1}$ on $(q,p]$, we refer to $\mathcal{Y}_n =\{ y_1,...,y_{n}\}$ as the set of \emph{sample curves}  and to $y_{n+1}$ as the \emph{focal curve}.

\subsection{Restricted and extended central regions}

Let $\mathcal{J}$ be a set with two or more sample curves and denote by $\mathcal{J}^+$ the set obtained by adding the focal curve to $\mathcal{J}$.

For each $y\in {\cal J}^+$, consider the modified band depth of \cite{pintadoromo2009} with bands formed by two curves
\begin{equation*}
\label{eq:MBD}
D_{[0,q]}(y,{\cal J}^+)=  \frac{1}{2}{m+1 \choose 2}^{-1}\sum_{x,z\in \mathcal{J}^+}  \lambda \left(\left\lbrace t\in [0,q]: \min (x(t),z(t)) \leq y(t) \leq \max (x(t),z(t)) \right\rbrace \right),
\end{equation*}
$\lambda(\{t\in [0,q]:A(t)\})$ being the proportion of time that $A(t)$ is true on $[0,q]$.
Roughly, $D_{[0,q]}(y,{\cal J}^+)$ is a measure of ``centrality'' or ``outlyingness'' of $y$ with respect to ${\cal J}^+$ on $[0,q]$.
%We leave for later a discussion about alternative measures of depth.
 
Denote by ${\cal J}_k$  the $k$ deepest curves of ${\cal J}$ according to $D_{[0,q]}(\cdot,{\cal J}^+)$.
This is, the $k$ curves of ${\cal J}$ with largest $D_{[0,q]}(\cdot,{\cal J}^+)$ values.
Note that the focal curve does not belong to ${\cal J}_k $ although it may be the deepest curve of ${\cal J}^+$.  
In this paper we consider the \emph{Restricted Central Region} (RCR) on $[0,q]$ delimited by the curves in ${\cal J}_k$  
\begin{equation}
\label{eq:centralregion}
R_{k}({\cal J}) = \{(t,y(t)): t\in [0,q], \min_{x\in {\cal J}_k}x(t) \leq y(t) \leq \max_{x\in {\cal J}_k} x(t)\}
\end{equation}
and its \emph{extension} on $[q,p]$ 
\begin{equation}
\label{eq:extendend}
\bar{R}_{k}({\cal J}) = \{(t,y(t)): t\in [q,p], \min_{x\in {\cal J}_k}x(t) \leq y(t) \leq \max_{x\in {\cal J}_k} x(t)\}.
\end{equation}

We remark RCRs differ slightly from ordinary central regions based on band depth \citep{pintadoromo2007}.
These regions can reveal features such as magnitude and shape of the curves in play \citep{sungenton2011}, particularly of the deep curves in $ {\cal J}^+$.
In principle, this utility is restricted to the interval where we measure  the depth, this is $[0,q]$. 
%In general, the extended regions do not have to be useful on  $[q,p]$.
However, if there is correlation between what is observed on $[0,q]$ and what is observed on $[q,p]$, one expects that the extended RCRs have properties on $[q,p]$ similar to those of RCR on $[0,q]$.
This is a key idea in our approach.
The correlation that we mention often arises from processes which are a mixture of randomness and periodicity, very common in stochastic modelling. This fact makes to our method a potential approach for a broad range of functional time series.

%Certainly, if  $y_{n+1}$ is not one of the $k$ deepest curves of ${\cal J}^+$ then $R_{k}({\cal J})$ corresponds to the $k/m$  central region of ${\cal J}$ discussed by of\cite{sungenton2011}.
%On the other hand, if $y_{n+1}$ is one of the $k$ deepest curves of ${\cal J}^+$ and is completely covered by $R_{k}({\cal J})$ then this RCR matches with the $(k+1)/m$  central region of ${\cal J}$.
%Otherwise, $R_{k}({\cal J}) $ does not have to coincide with any ordinary central region.

%For simplicity, in this paper we only consider the modified depth measure. \cite{sungenton2011},
%However, other depth measures could be used to develop alternatives methods of forecasting.
%In fact, we considered the centraegions used by \cite{Hyndman2010}. 
%We refer the interested reader to the Appendix, where the results obtained from this alternative are presented as illustration.

\subsection{A depth-based algorithm for focal-curve enveloping}
\label{sec:curvesSelection}

From the above, given a focal curve $y_{n+1}$  and a set of sample curves  $ {\cal Y}_n$, we are interested on subsets of sample curves ${\cal J}$ such that $y_{n+1}$ is deep in ${\cal J}^+$, seeking that the shape and magnitude of $y_{n+1}$ be captured by RCRs of ${\cal J}$.
Of the $2^n-1-n$ possible sets of sample curves, many of them may have RCRs which scarcely covers the focal curve, even if they are nearby (here we use Euclidean distance for measuring nearness but the method can be straightforwardly adapted to other distances). 
In general, the nearest curves to $y_{n+1}$ may not envelope it.
Conversely, many  subsets may completely envelope the focal curve but with wide RCRs, whose boundaries are faraway from $y_{n+1}$. 
These regions do not provide useful information about the features of the focal curve.

\begin{figure}[]
\begin{center}
\includegraphics[width=5cm, height=7cm]{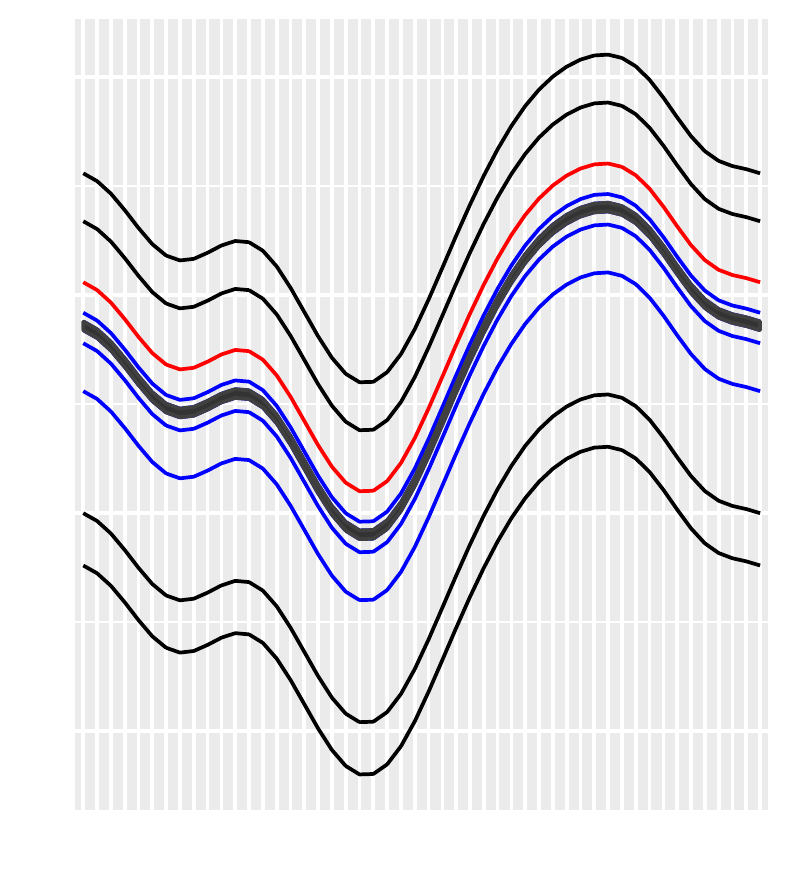}
\includegraphics[width=5cm, height=7cm]{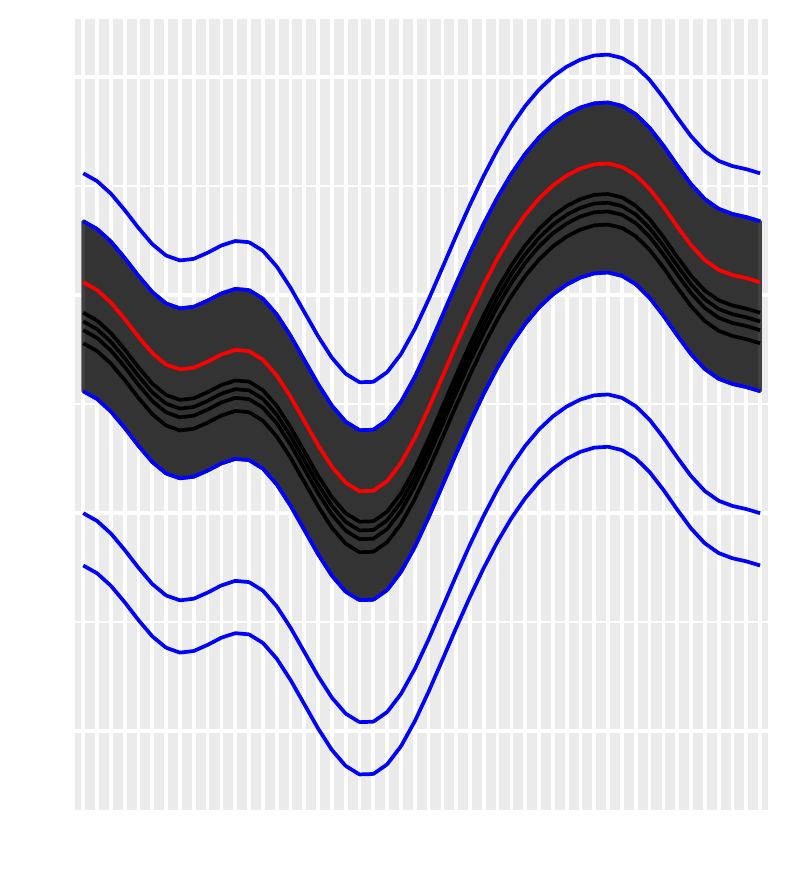}
\includegraphics[width=5cm, height=7cm]{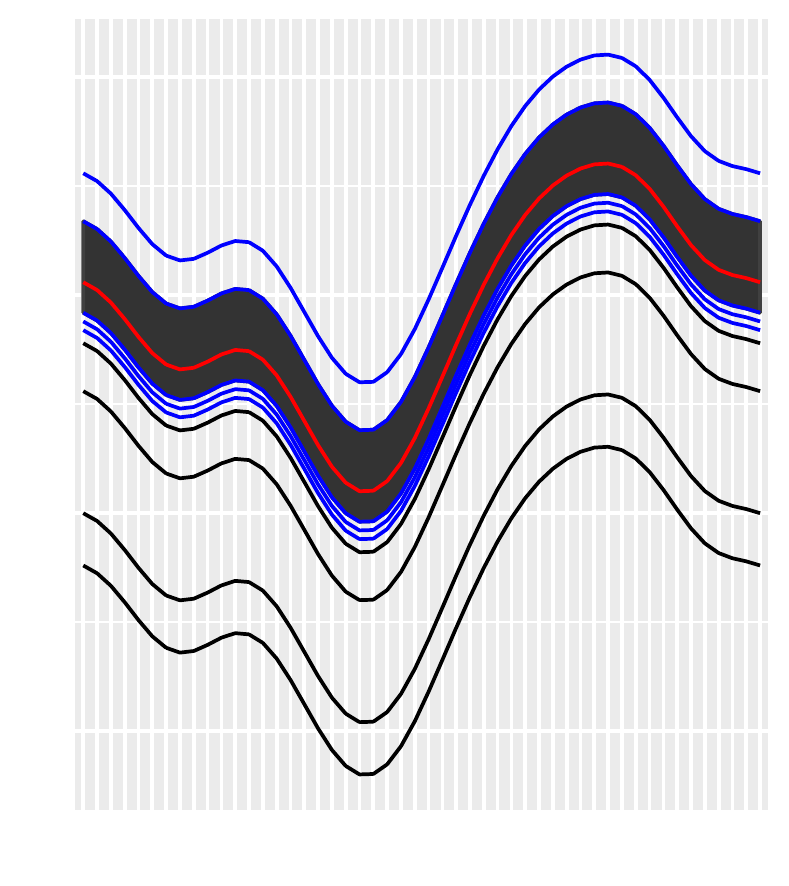}
\caption{Three scenarios of bands (in gray) delimited by the two deepest curves of five blue curves.
The blue curves correspond to (left panel) the five nearest lines to the red curve, (central panel) the five farthest curves, and (right panel) the set with tightest gray band that envelopes the focal curve. }
\label{fig:illustration}
\end{center}
\end{figure}

To illustrate the argument above, consider the ten curves shown in Figure~\ref{fig:illustration}.
Three different subsets ${\cal J}$ of five blue curves are considered.
The RCRs delimited by the two deepest curves of ${\cal J}$ are shown in grey.
At the left panel, ${\cal J}$ is composed with the five nearest curves to the red one.
Whereas at the central panel, ${\cal J}$ consists of the five farthest curves.
Our goal is to compute RCRs as the one shown in right panel.
This is the tightest RCR that envelopes the focal curve.

\begin{figure}[]
\begin{center}
\includegraphics[width=16cm, height=7cm]{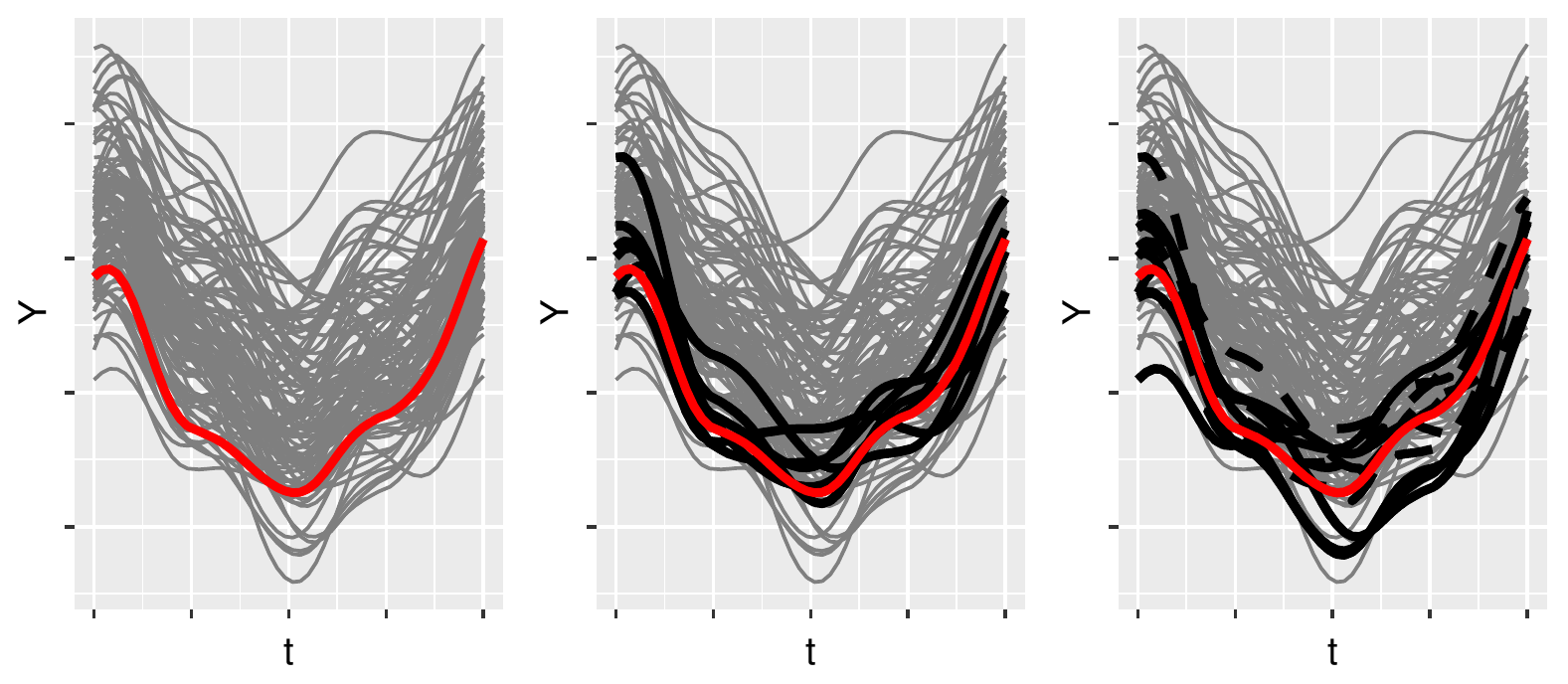}
\caption{Left panel: sample curves (grey) and focal curve (red). Center panel: curves selected at the first iteration (black). Right panel: curves selected at the second/last iteration (black) and first iteration (dashed lines). }
\label{fig:Alg2}
\end{center}
\end{figure}

We will address the problem by selecting curves \emph{from $y_{n+1}$ to outwards}, 
enveloping and making $y_{n+1}$ a deep datum.
For this, first we identify the set $I_q\subset [0,q]$ where $y_{n+1}$ is enveloped by the sample. 
This is 
\begin{equation}
I_q = \{ t \in [0,q] : \min_{y \in {\cal Y}_n} y(t) \leq y_{n+1}(t) \leq \max_{y \in {\cal Y}_n} y(t) \}.
\end{equation}
We assume $I_q$ is not empty and review curves, from the nearest curve to the farthest from $y_{n+1}$, to select those that contribute to cover $y_{n+1}$  until this is completely enveloped on $I_q$.
We gather these curves in ${\cal J}$.
Next, we repeat the process but using the curves in ${\cal Y}_n\setminus {\cal J}$  and save temporally the selected curves in ${\cal N}$.
If the depth of $y_{n+1}$ in ${\cal J}^+\cup {\cal N}$ does not decrease  in relative terms respect to the depth in ${\cal J}^+$ then we add ${\cal N}$ to ${\cal J}$, otherwise we remove ${\cal N}$ from ${\cal Y}_n$, and repeat the last process until there are not curves to select.
We refer to the set of curves collected in ${\cal J}$  as the \emph{focal-curve envelope}.
Properties of ${\cal J}$ are:
by construction, the band delimited by the curves in ${\cal J}$ covers the focal curve on $[0,q]$ as much as it is possible to do with the whole sample.
Second, the nearest curves to $y_ {n + 1}$ often belong to ${\cal J}$, although this is not usually a set of ``$k$-nearest curves''.
Finally, for the cases in which there exists a set of curves  such that the focal curve is the deepest on $[0,q]$, we have observed that  $y_{n+1}$ is also the deepest of ${\cal J}^+$. 
As illustration, Figure~\ref{fig:Alg2} shows first and last (second) iteration of the algorithm sketched above by considering 100 curves from a simulated functional time series. 
Note how the curves selected on the first iteration are near and cover the focal curve while the curves selected in the next iteration make it be the deepest curve of the envelope.
To avoid ambiguities about how the algorithm works, we outline its steps  in  pseudocode (see Appendix).

\subsection{Forecasting method}

For band forecast, we consider extended RCRs of the focal-curve envelope $\mathcal{J}$, equation~\ref{eq:extendend}. 
Two statistics that either separately or combined are always considered for evaluating band forecast are: On one hand, the \emph{coverage}; in our case, the proportion of time that the focal curve is in $\bar{R}_{k}({\cal J})$. %also called empirical coverage probability \citep{hyndmanshang2011}. 
This is,
\begin{equation}
C_k({\cal J})= \lambda(\{t\in [q,p]:(t,y_{n+1}(t))\in \bar{R}_{k}({\cal J})\}).
\end{equation}
On the other hand, the \emph{band mean width}, that we standardize for comparing results on different functional time series. Namely,
\begin{equation} 
W_k({\cal J}) = \sum_{t\in [q,p]} \left(\max_{y\in {\cal J}_k} y(t) -\min_{y\in {\cal J}_k}y(t)\right) /  \sum_{t\in [q,p]} \left(\max_{y\in {\cal Y}_n} y(t) -\min_{y\in {\cal Y}_n}y(t)\right).
\end{equation} 
This is a measure to assess how narrow is a prediction band relative to the band delimited by the whole sample.
High coverage and small mean width are desirable to capture magnitude and shape of the focal curve.
But both $C_k$ and $W_k$ are nondecreasing on $k$, therefore the selection of $k$ involves a trade-off between coverage and mean width.
The decision problem may be addressed by the graphical tool that we describe below.

Unlike mean width, coverage is random, it depends on the unobserved part of the focal curve.
Hence, we consider expected coverage versus mean width for tuning $k$.
Let $\mu_k = \mathbb{E}[C_k({\cal J})]$.
However, high expected coverage does not guarantee at all high coverage, not even high coverage with high probability. 
For this reason, for preventing poor coverages, we also consider a bottom threshold for the coverage with a $(1-\alpha)\times 100\%$ confidence level.
Namely, the $\alpha\times 100\%$ percentile of the probability distribution of $C_k(\mathcal{J})$, denoted here by $c_k^\alpha$.
This is $\mathbb{P}(C_k({\cal J})\leq c_k^\alpha)=\alpha$.
The idea, then, is to select $k$ by taking into account $\mu_k$,  $c_k^\alpha$ and $W_k$.
In practice, $\mu_k$ and $c_k^\alpha$ must be estimated.
For this, we consider the following plug-in approach:

Consider the $m$ most recent periods previous to the focal one.
%For example, if periods are days, and the full sample cover several years m should be 365.
Then consider the $m$ envelopes $\mathcal{J}(i)$, $n-m+1\leq i\leq n$, obtained by restricting data of $y_{i}$ to $[0,q]$ and computing the focal-curve envelope of $y_{i}$ from the sample curves $y_{1},...,y_{i-1}$.
The average and the $\alpha\times 100\%$ percentile of observed coverages $C_k(\mathcal{J}(n-m+1)),\dots,C_k(\mathcal{J}(n))$, that we denoted by $M_k$ and $C_k^\alpha$, are natural estimators of $\mu_k$ and $c_k^\alpha$. 
The underlying idea is that under ergodic hypothesis about the functional time series, time averages should be similar to expected values.
For this, we must assume that $m$ and $n-m$ are large.
We explore this conjecture  by simulation in next section. 
In particular, we provide statistical evidences for conjecturing $M_k$ is an unbiased estimator of $\mu_k$ and $\mathbb{P}(C_k({\cal J})\geq C_k^\alpha)\geq1-\alpha$.

\begin{figure}[]
\begin{center}
\includegraphics[width=15cm]{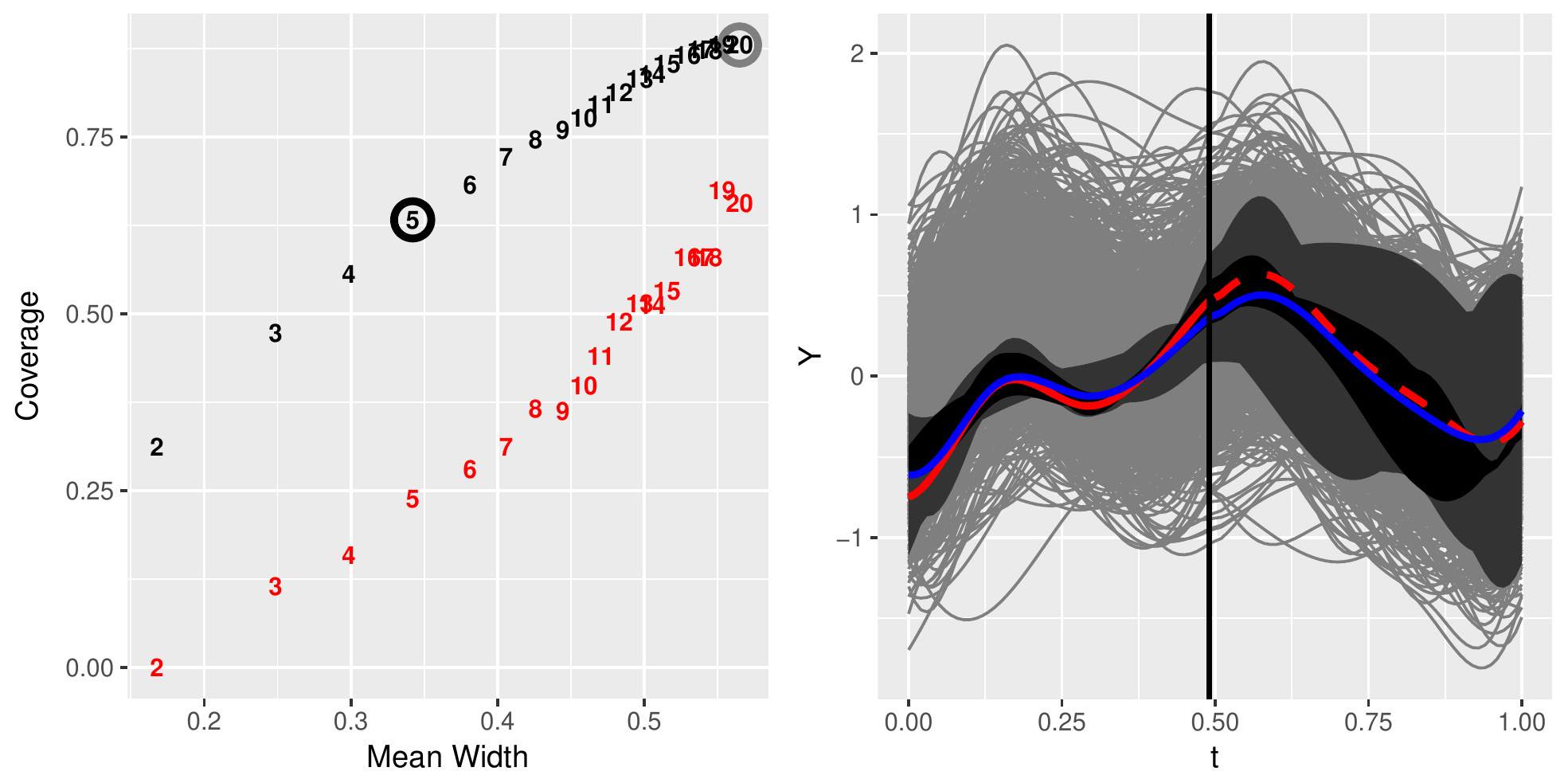}
\end{center}
\caption{Left panel: Estimated expected coverage (black) for different $k$-values and corresponding lower bounds for a confidence level of 90\% and mean widths. Right panel: Sample curves (grey), focal curve (red, observed part in solid line),  restricted and extended central regions for $k=5$ and 20 (black and grey, respectively), and weighted functional mean of the focal-curve envelope (blue).}
\label{fig:simulPractitioner}
\end{figure}

Given a confidence level of $(1-\alpha)\times 100\%$, we plot $(W_k(\mathcal{J}), C_k^\alpha)$ and $(W_k(\mathcal{J}), M_k)$ for several possible values of $k$.
From this plot, the practitioner should select a $k$ value that fits her/his preferences about mean width, expected and minimum coverage for a given confidence level.
As illustration, we plot the resulting chart  (left panel of Figure~\ref{fig:simulPractitioner}) by considering a simulated functional time series  divided into $n=1000$ sample curves and a focal curve with $p=1$, $q=1/2$, $m=100$ and $\alpha = 0.1$.
We also show (right panel) two RCRs and its extensions corresponding to $k = 5$ and $20$.
From this plot we expect a coverage on the prediction interval above 62.5\% for $k=5$ and 87.5\% for $k= 20$.
Also, with a confidence level of $90\%$, the coverage will be larger than 25\%  for $k=25$ and larger than 62.5\% for $k= 20$.

For point forecast, we propose the weighted functional mean of the focal-curve envelope $\mathcal{J}$. This is,
\begin{equation}
\hat{y}_{n+1} = \sum_{y\in {\cal J}}  w_y y,\ \ \mbox{with} \ \ w_y = d(y, y_{n+1}) \large/\sum_{y\in {\cal J}} d(y, y_{n+1}),
\end{equation}
 $d(y, y_{n+1})$ being the Euclidean distance between $y$ and $y_{n+1}$ on $[0,q]$.
Figure \ref{fig:simulPractitioner} (right panel) shows also point forecast for the example considered above.

%The nominal probability $1-\alpha$ can be compared with the empirical probability, calculated by repeating the method on several focal periods.
% let us say from $n+1$ to $n+m$. This is
%\begin{equation}
%p_\alpha = \frac{1}{m} \sum_{i=n+1}^{n+m}\mathbbm{1}_{\{C_{k(i-1,\alpha)}(\mathcal{J}(i))\geq c\}},
%\end{equation}
%$\mathbbm{1}_A$ being the indicator function of $A$. 

\section{Simulation Study}

%We compare estimators and observed values from different data sets.
%We cover feasible ranges of $k$, no matter if some RCRs have low coverages or high mean widths.
%The proper selection of the $k$ values is a decision of the practitioner who must consider the trade-off between coverage and width.
%The purpose of the comparative study is to assess the performance of the statistical tools that we provide and not how to used them.
%For this, we show for each test set a particular example of using.

%For evaluating the method, we split data into natural cycles of $m$ periods and use the more recent cycle for testing.
%On the testing cycle we compute mean square error (MSE), i.e. the  Euclidean distance between the focal curve and its point estimator on $[q,p]$.
%We also compute empirical probabilities, by applying the method on testing cycle, and compare them with their nominal values.

%\subsection{Periodically correlated processes }

The class of the periodically correlated (PC) processes sets up a framework for modeling functional time series with complex periodic rhythm.
%They have been used as models in many fields, such that meteorology, radio physics, communications engineering and finance.
They are processes whose mean and autocovariance function are periodic functions with same period.
These models were introduced by \cite{gladyshev1961}, since then they have been subject of study by several authors.
The  book of \cite{harry2007} presents the main theory as well as applications to meteorology, climate, communications, economics, and machine diagnostics.

%
%This has sense when the shape and magnitud of $y_0$ is captured by the curves  of $\mathcal{J}_k(y_0)$, not only on $I$ but on $I_0$.
%
%if the seasonal factors have a significant influence on the series, we expect that the shape and magnitud of $y_0$ is captured by the deepest curves  of $\mathcal{J}(y_0)$, not only on $I$ but on $I_0$.
%
%We remark that the deepest curves of $\mathcal{J}(y_0)$ are nearby curves that surround $y_0$ on $I$.
%If the seasonal factors have a significant influence on the series, we expect that the shape and magnitud of $y_0$ is captured by the deepest curves  of $\mathcal{J}(y_0)$, not only on $I$ but on $I_0$.
%%
%This could happen if the process is correlated and its trajectories show periodic patterns.
%In this case, it seems intuitively clear that if $y_0$ is enveloped on $I$ by a set of near curves then they also will surround $y_0$ on $I_0$.
%
%Next, we explore this idea by simulation.

We consider two stationary Gaussian processes that we combine to create a wide palette of PC processes.
First, a Gaussian process $X$ with zero mean and squared exponential autocovariance function
 \begin{equation}
\mbox{Cov}(X(t),X(s)) = \sigma_X^2\exp(-|t-s|^2/2\textit{l}_X^2).
\end{equation}
%\textit{l} being the characteristic length-scale of the process.
This is the default autocovariance function in Gaussian processes simulation \citep{rasmussen2005}.
The \emph{lengthscale}  $\textit{l}_X$  determines the length of the `wiggles' in the trajectories.
While $\sigma_X$ determines the average distance of the trajectories away from zero.
Second, a Gaussian process $f$, independent of $X$, with zero mean and periodic covariance function,
\begin{equation}
\mbox{Cov}(f(t),f(s)) =  \sigma_f^2 \exp(-2\sin^2(\pi |t-s|/p)/\textit{l}_f^2).
\end{equation}
The parameters $\textit{l}_f$  and $\sigma_f$ determine lengthscale and  average distance in the same way as in the squared exponential autocovariance function.
We emphasize that the trajectories of $f$ are periodic functions of period $p$.

\begin{figure}[]
\begin{center}
\includegraphics[height=11cm]{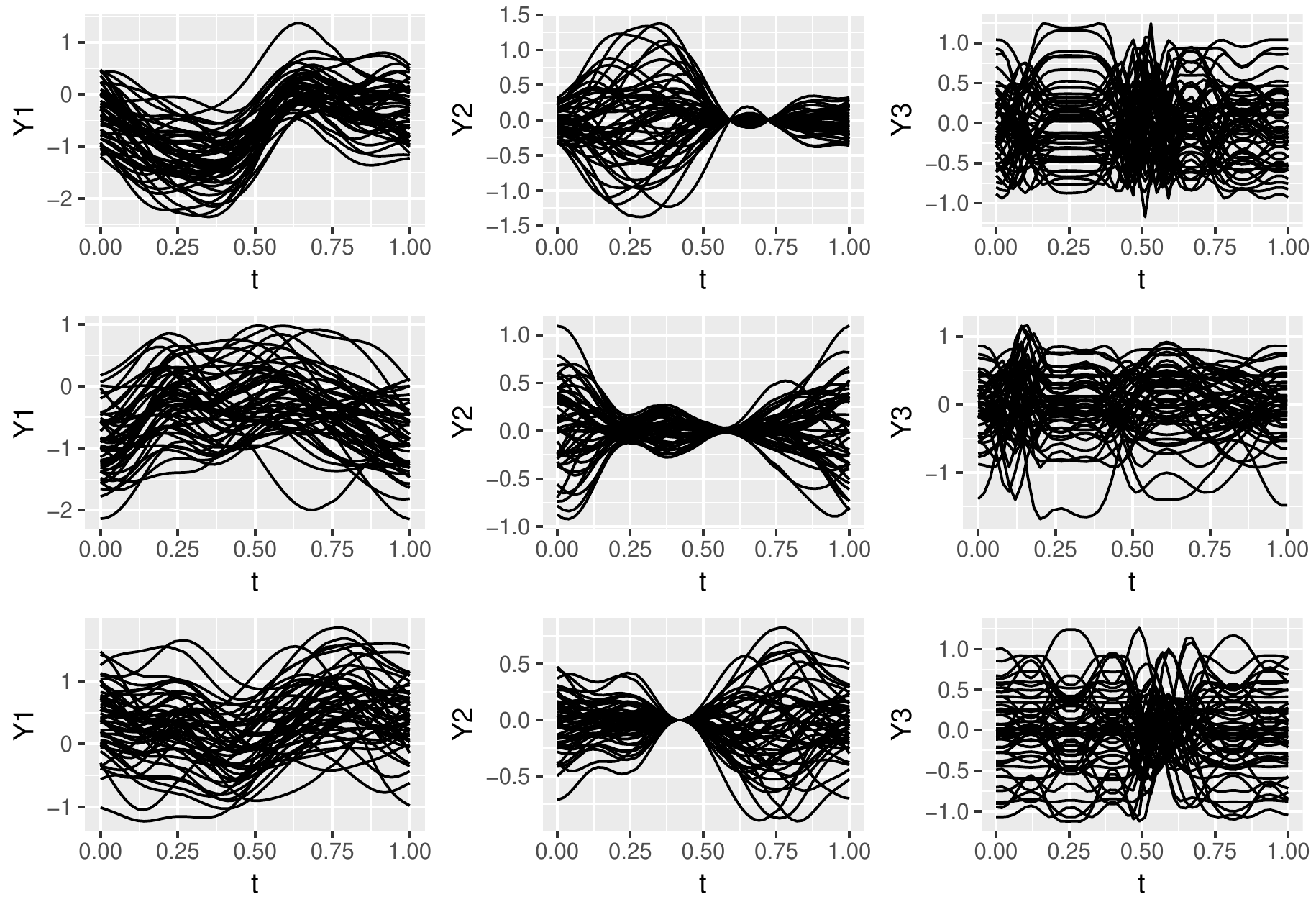}
\end{center}
\caption{Examples of sample curves by slicing nine different trajectories of $Y_1, Y_2$ and $Y_3$. The same observed periodic component from $f$ is used by row and the same observed irregular trajectory from $X$ is used by columns.}
\label{fig:examplepatterns}
\end{figure}

We consider three types of seasonal signals:
\begin{equation}
Y_1(t) = f(t) + X(t), \  Y_2(t) = f(t)  X(t), \ \mbox{and} \ Y_3(t) = X(t +f(t)).
\end{equation}
Sums of a seasonal and  irregular component, as $Y_1$, have been widely used.
In particular, $Y_1$ is stationary, although its trajectories exhibit a periodic pattern of length $p$.
However, $Y_2$ and $Y_3$ are non-stationary, they are periodically correlated with period $p$. % (see Supplementary Material).
If $f$ were deterministic then $Y_2$ would correspond to an \emph{amplitude modulation} of a stationary process while $Y_3$ to a \emph{time-scale modulation}, frequently used signals in engineering. 
We randomize $f$ by considering a Gaussian process with periodic covariance function for producing a wide spectrum of sample curves.
Figure~\ref{fig:examplepatterns} shows some examples of curves based on three observed patterns of $f$, with $p=1$, $\sigma_X/\sigma_f = 1$ and $\textit{l}_X/\textit{l}_f = 0.2$.

\subsection{Testing confidence level and expected coverage estimation}

Before testing the goodness of our forecasting method, we test the two basic properties on which the practitioner bases her/his selection about which prediction band to use.
They are, $\mathbb{E}[M_k]=\mu_k$ and $\mathbb{P}(C_k({\cal J})\geq C_k^\alpha)\geq1-\alpha$.
We address this problem by using Monte Carlo. 

\begin{figure}[]
\begin{center}
\includegraphics[width=15cm]{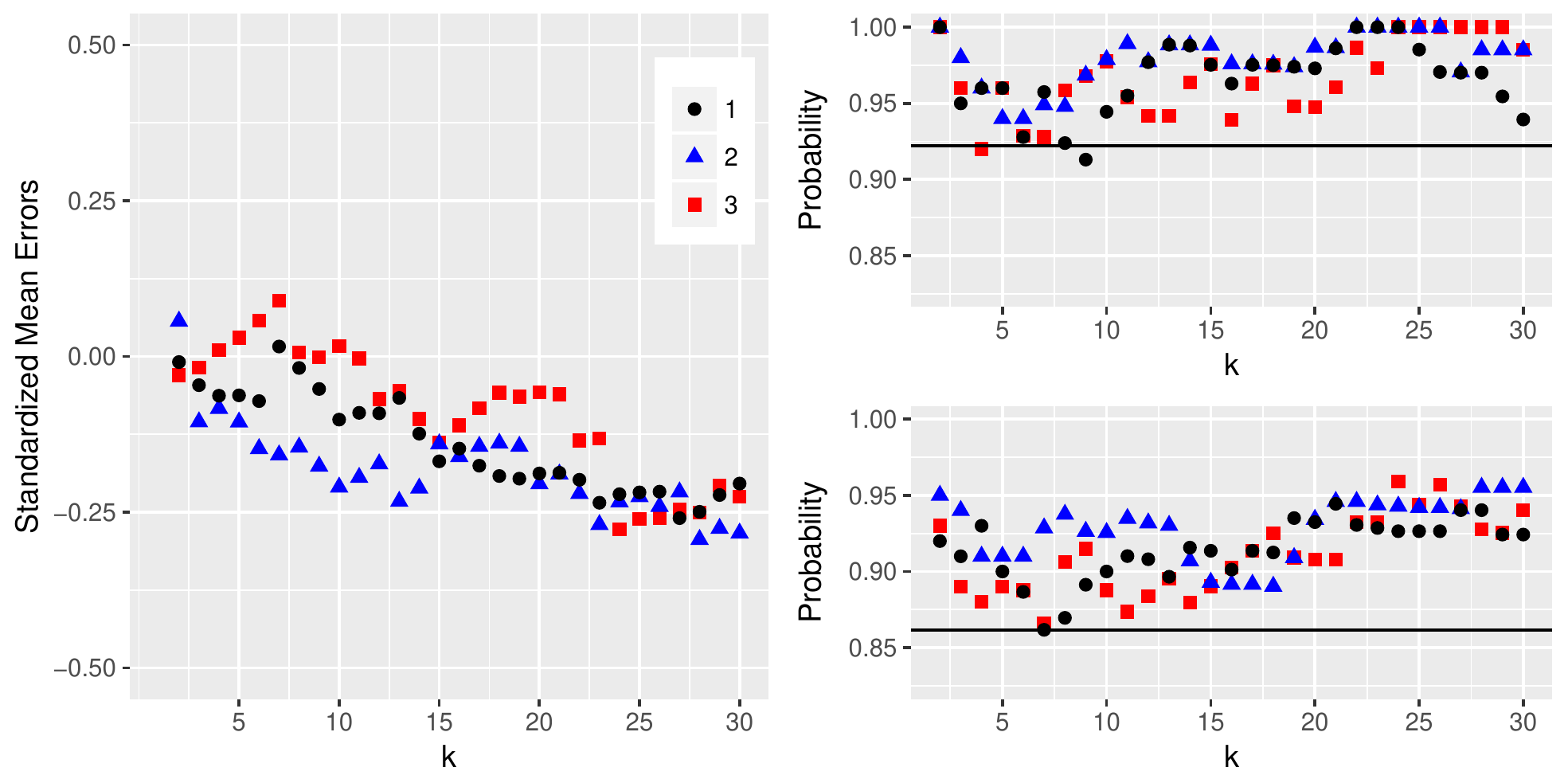}
\end{center}
\caption{Left panel: standardized mean error of mean coverage estimators based on $100$ independent trajectories of  $Y_i$, $i=1$, $2$ and $3$. Right panel: Empirical probabilities that the observed coverage is greater than the bottom threshold proposed, for $\alpha = 0.1$ (bottom) and $\alpha=0.05$ (top). Black lines delimite rejection regions at $95\%$ confidence level for the hypothesis: the above probability is $\geq$ than $1-\alpha$.}
\label{fig:testingproperties}
\end{figure}

We simulated $N$ independent trajectories of $Y_i$, $i=1, 2$ and $3$.
Each trajectory with 1001 periods, making $n=1000$ and reserving the last period as focal, with $q=1/2$.
We fixed $m=100$ and observed $M_k$ and $C_k({\cal J})$ for each trajectory, obtaining a paired random sample of size $N$ of these statistics.
Let $d_j^k$ be the observed value of $M_k - C_k({\cal J})$ for the $j$th trajectory.
Then, under the null hypothesis $H_0: \mathbb{E}[M_k] = \mu_k$, the Central Limit Theorem implies that the standardized mean error $\sqrt{N} \times \mbox{mean}(\{d_j^k\}_{j=1}^N)/\mbox{std}(\{d_j^k\}_{j=1}^N))$ is approximately standard normal when $N$ is large.
We computed this error for a wide range of $k$ values and $N=100$ and observed all of them fallen between $-0.3$ and $+0.3$ (see left panel of Figure \ref{fig:testingproperties}).
Roughly, we have strong evidences against an alternative to $H_0$.
A similar asymptotic argument based on the proportion of trajectories for which  $C_k({\cal J}) \geq C_k^\alpha$ (i.e., the empirical probability) holds to reject the hypothesis  $\mathbb{P}(C_k({\cal J}) \geq C_k^\alpha) < 1-\alpha$.
Results for $\alpha = 0.1$ and $0.05$ are shown in right panel of Figure \ref{fig:testingproperties}.
In conclusion, the graphical tool for addressing the problem of choosing $k$ works as we expected, at least for the considered PC processes.
%
%d a focal slice and considered $I=[0,p/2]$ and $I_0=(p/2,p]$.
%For illustration, we show in Figure~\ref{fig:sinesBandExample} bands with $k=6$ for a trajectory of $Y_1, Y_2$ and $Y_3$.
%The first thing one can notice is that the shape and magnitude of the focal is captured by the bands on both intervals, $[0,p/2]$ and $(p/2, p]$, no matter which algorithm we are using.
%This is confirmed by the results of our simulation.
%Table~\ref{tab:simulation} shows that although the depth of the focal decreases in the prediction interval, it remains significantly deep (among the first third) and the curve is covered similarly in both intervals $[0,p/2]$ and $(p/2, p]$.
%We also observe that the total number of curves chosen of both algorithms is roughly the same.
%In Figure~\ref{fig:CWSINES}, we present averages and standard deviation of coverage and width for the prediction interval $(p/2,p]$.
%This figure shows that bands delimited by the $k$ deepest curves cover the focal one on a large proportion, providing tight prediction bands for proper selection of $k$. 

\subsection{Forecasting exercises with PC processes}

We consider three scenarios for testing our forecasting method. 
First, a demanding practitioner with high coverage expectation (90\%) and high threshold of minimum coverage (60\%) without taking into account band mean width. 
Therefore, this practitioner looks for the smallest $k$ such that $M_k \geq 0.9$ and $C_k^\alpha \geq 0.6$. 
Although it is not the case for the simulations that we report below, we remark that it could happen that the focal-curve envelope has not enough sample curves for satisfying such condition on $k$. 
Second, a conservative practitioner with high coverage expectation but low threshold of minimum coverage (30\%).
This involves the smallest $k$ such that $M_k \geq 0.9$ and $C_k^\alpha \geq 0.3$. 
Finally, a conformist practitioner who is satisfied with $M_k \geq 0.6$ and $C_k^\alpha \geq 0.3$. 

In table~\ref{tab:simulResults}, we observe the results when the three practitioner described above forecast the second half of the last 100 periods of one trajectory with 1,001 recorded periods of model $Y_i$, $i=1, 2$ and $3$.
For these exercises we consider $\alpha = 0.05$ and $0.10$.
The average of the selected $k$ (that we denote by $k^*$) across the 100 trials  and the averages of the resulting coverage $C_{k^*}({\cal J})$ and mean width $W_{k^*}({\cal J})$ are reported in Table 1.
It is also reported the empirical probability of $C_{k^*}({\cal J}) \geq 0.6$ for the first practitioner and the corresponding proportion of $C_{k^*}({\cal J}) \geq 0.3$ for the other two.
Remarkably, the empirical probability fits the nominal one (0.90 and 0.95).
This is, the proposed threshold of minimum coverage works for the two confidence levels considered.
On the other hand, the observed coverage is larger in average than the expected coverage.
This is because the selection rule of $k$ typically involves larger $k$-values than those required to satisfy only a coverage-expectation condition.
Finally, we remark that Practitioner 3 obtains high coverages and low mean width.

\begin{table}[]
\centering
\begin{tabular}{lllllllll}
\cline{2-9}
                              & \multicolumn{4}{c|}{$\alpha = 0.05 $}                                                 & \multicolumn{4}{c}{$\alpha = 0.10 $}                       \\ \cline{1-9} 
                              & \multicolumn{8}{c}{$Y_1$}                                                                                                                          \\ \hline
\multicolumn{1}{l|}{Averages} & $C_{k^*}({\cal J})$ & $W_{k^*}({\cal J})$       & Prob. & \multicolumn{1}{l|}{$k^*$}  & $C_{k^*}({\cal J})$ & $W_{k^*}({\cal J})$ & Prob. & $k^*$  \\
\multicolumn{1}{l|}{Pract. 1} & 0.941               & \multicolumn{1}{c}{0.511} & 0.949 & \multicolumn{1}{l|}{20.913} & 0.913               & 0.504               & 0.895 & 19.041 \\
\multicolumn{1}{l|}{Pract. 2} & 0.938               & 0.509                     & 0.966 & \multicolumn{1}{l|}{20.588} & 0.876               & 0.468               & 0.895 & 14.570 \\
\multicolumn{1}{l|}{Pract. 3} & 0.758               & 0.377                     & 0.935 & \multicolumn{1}{l|}{8.060}  & 0.707               & 0.332               & 0.918 & 5.800  \\ \hline
\end{tabular}

\begin{tabular}{lllllllll}
                              & \multicolumn{8}{c}{$Y_2$}                                                                                                                          \\ \hline
\multicolumn{1}{l|}{Averages} & $C_{k^*}({\cal J})$ & $W_{k^*}({\cal J})$       & Prob. & \multicolumn{1}{l|}{$k^*$}  & $C_{k^*}({\cal J})$ & $W_{k^*}({\cal J})$ & Prob. & $k^*$  \\
\multicolumn{1}{l|}{Pract. 1} & 0.942               & \multicolumn{1}{c}{0.529} & 0.952 & \multicolumn{1}{l|}{21.256} & 0.929               & 0.509               & 0.926 & 18.878 \\
\multicolumn{1}{l|}{Pract. 2} & 0.940               & 0.528                     & 0.952 & \multicolumn{1}{l|}{20.870} & 0.902               & 0.478               & 0.894 & 15.060 \\
\multicolumn{1}{l|}{Pract. 3} & 0.807               & 0.388                     & 0.948 & \multicolumn{1}{l|}{8.170}  & 0.731               & 0.329               & 0.910 & 5.780  \\ \hline
\end{tabular}

\begin{tabular}{lllllllll}
                              & \multicolumn{8}{c}{$Y_3$}                                                                                                                          \\ \hline
\multicolumn{1}{l|}{Average}  & $C_{k^*}({\cal J})$ & $W_{k^*}({\cal J})$       & Prob. & \multicolumn{1}{l|}{$k^*$}  & $C_{k^*}({\cal J})$ & $W_{k^*}({\cal J})$ & Prob. & $k^*$  \\
\multicolumn{1}{l|}{Pract. 1} & 0.915               & \multicolumn{1}{c}{0.461} & 0.958 & \multicolumn{1}{l|}{19.263} & 0.915               & 0.455               & 0.900 & 17.700 \\
\multicolumn{1}{l|}{Pract. 2} & 0.901               & 0.451                     & 0.931 & \multicolumn{1}{l|}{18.252} & 0.865               & 0.413               & 0.871 & 12.940 \\
\multicolumn{1}{l|}{Pract. 3} & 0.742               & 0.326                     & 0.927 & \multicolumn{1}{l|}{7.170}  & 0.662               & 0.289               & 0.860 & 5.370  \\ \hline
\end{tabular}

\caption{Three result sets obtained by predicting the second half of the last 100 periods of one trajectory with 1,001 periods.
Each trajectory corresponds to one of the three PC processes denoted $Y_1, Y_2$ and $Y_3$.
Practitioner 1 select the smallest $k$ such that $M_k \geq 0.9$ and $C_k^\alpha \geq 0.6$. Practitioner 2, the smallest $k$ such that  $M_k \geq 0.9$ but $C_k^\alpha \geq 0.3$. And. practitioner 3,  the smallest $k$ such that $M_k \geq 0.6$ and $C_k^\alpha \geq 0.3$.
Averages of observed coverage, mean width, empirical probability and selected $ k $ by the practitioner are grouped according to the confidence level used.
We remark that the empirical probability fits the nominal probability 0.90 and 0.95, respectively.}
\label{tab:simulResults}

\end{table}

\begin{figure}[]
\begin{center}
\includegraphics[height=6cm]{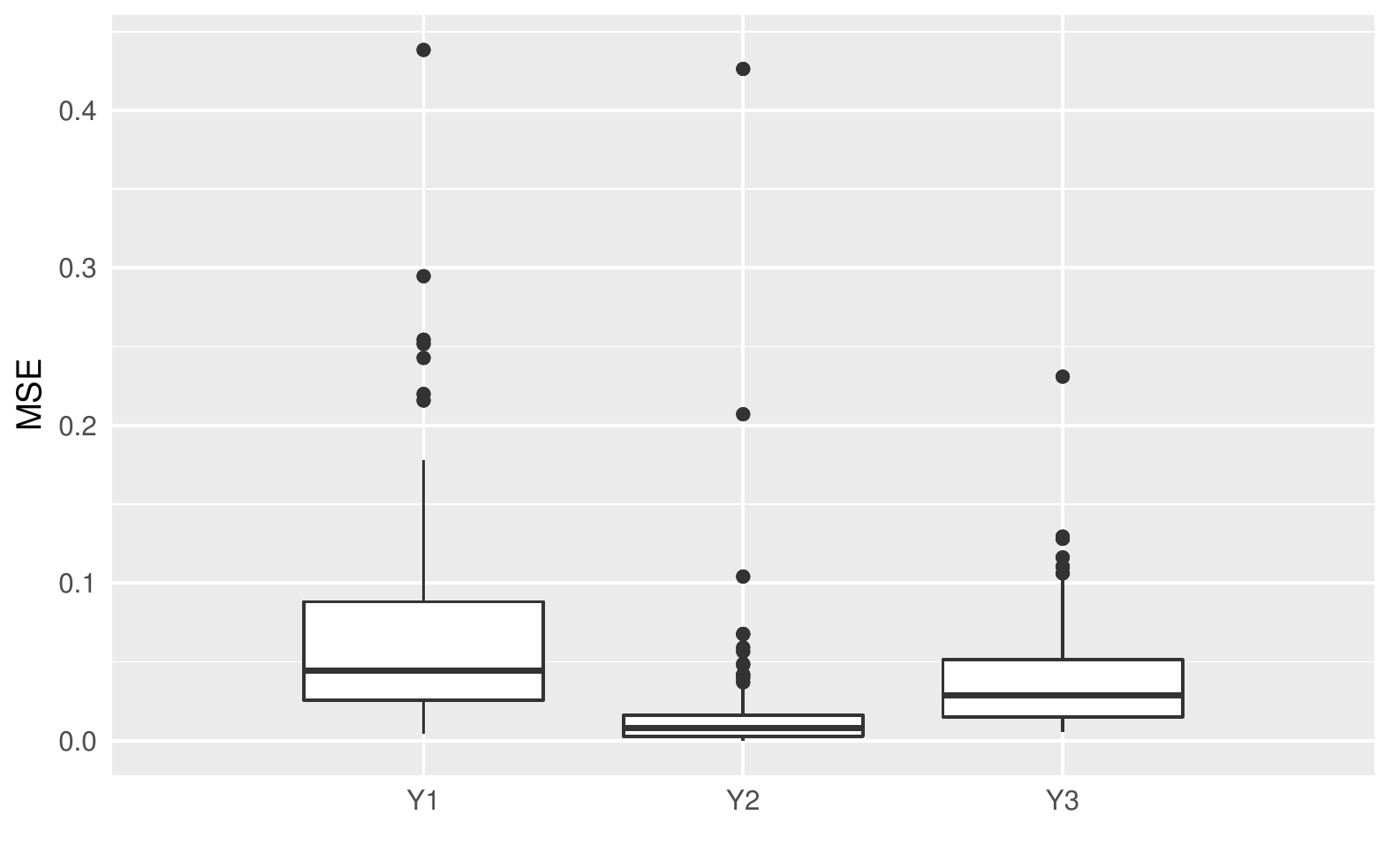}
\end{center}
\caption{Mean Square Errors of proposed point prediction from the forecasting exercise based on trajectories of PC processes $Y_1, Y_2$ and $Y_3$.}
\label{fig:MSE}
\end{figure}

\begin{figure}[]
\begin{center}
\includegraphics[width=16cm, height=5cm]{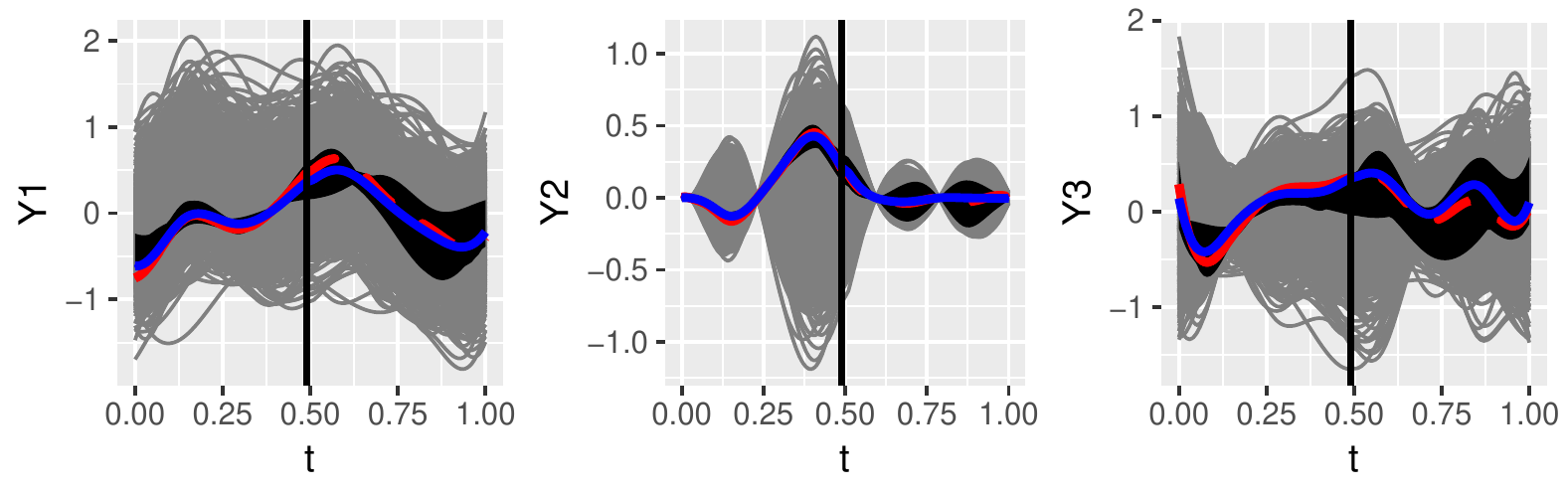}
\end{center}
\caption{Left-half side of panels: focal curve  (solid red line) and band (in black) based on the six deepest slices on $[0,1/2]$ provided by Algorithm 2. Right-half side of panels: extended extended bands and focal (dashed red line).}
\label{fig:sinesBandExample}
\end{figure}

Boxplots of mean square errors of point forecast are plotting in Figure~\ref{fig:MSE}, showing accurate prediction for the three considered models of PC processes.
As illustration, we also show prediction bands for the last period of each model in Figure~\ref{fig:sinesBandExample}.

\label{sec:cases}
\section{Case Study: Spanish electricity demand}

Data concerning to the Spanish electricity demand is available at \url{http://www.ree.es/es/},
from where we obtained the demand in megawatts (MW) from January first 2014 to December 31st  2017 each 10 minutes. 
Thus, we consider the daily demand, $1461$ curves in total.
The demand varies significantly during the day, setting certain daily pattern that we can directly observe from simple plots showed below.
Also it is reasonable to expect that the shape of the beginning of the day is correlated to the shape of the end, suggesting common characteristics with the PC processes discussed above.
Of course, demand is more complicated than PC processes, we remark ruptures from working days to weekends or holidays, not to mention the difference among seasons.
Otherwise, we could address the prediction problem estimating directly the PC structure.

Functional methods for electricity demand forecasting have been widely used
\citep{Vilar2012,Antoch2010,Paparoditis2013,Cho2013,Shang2013,Aneiros2013}.
We recommend  the paper of \cite{aneiros2016}, who also consider Spanish demand, for finding out more about functional data analysis for electricity forecasting.
The methods developed by these authors are based on regression techniques that relies on the chronological order of the days, being able to provide accurate prediction for next-day electricity demand.
Our functional data approach applied to this important problem is totally different and complementary. 
On one hand, we consider the practical case in which the demand of part of the day has been already observed and should be used for updating the prediction for the rest of the day.
On the other hand, we review past days, without take care about how recent they are, to capture the phenomenology of the day to predict with the only goal of providing forecast bands that may be useful to anticipate critical scenarios.

\begin{figure}[]
\begin{center}
\includegraphics[height=8cm, width=5cm]{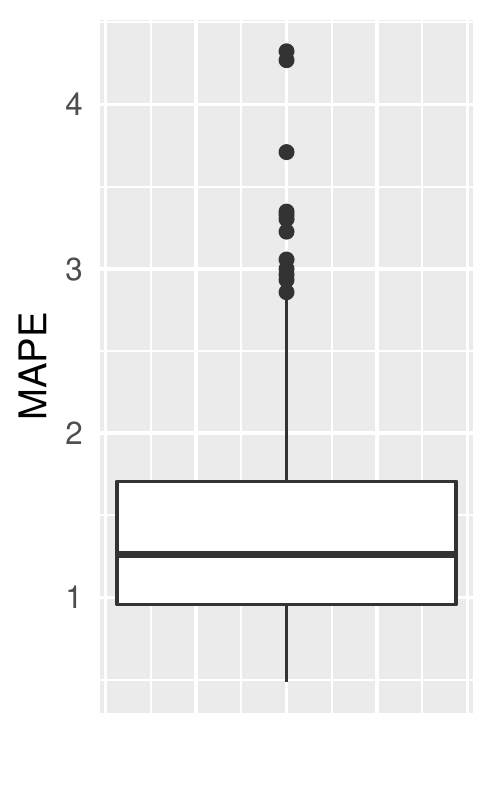}
\includegraphics[height=8cm, width=11cm]{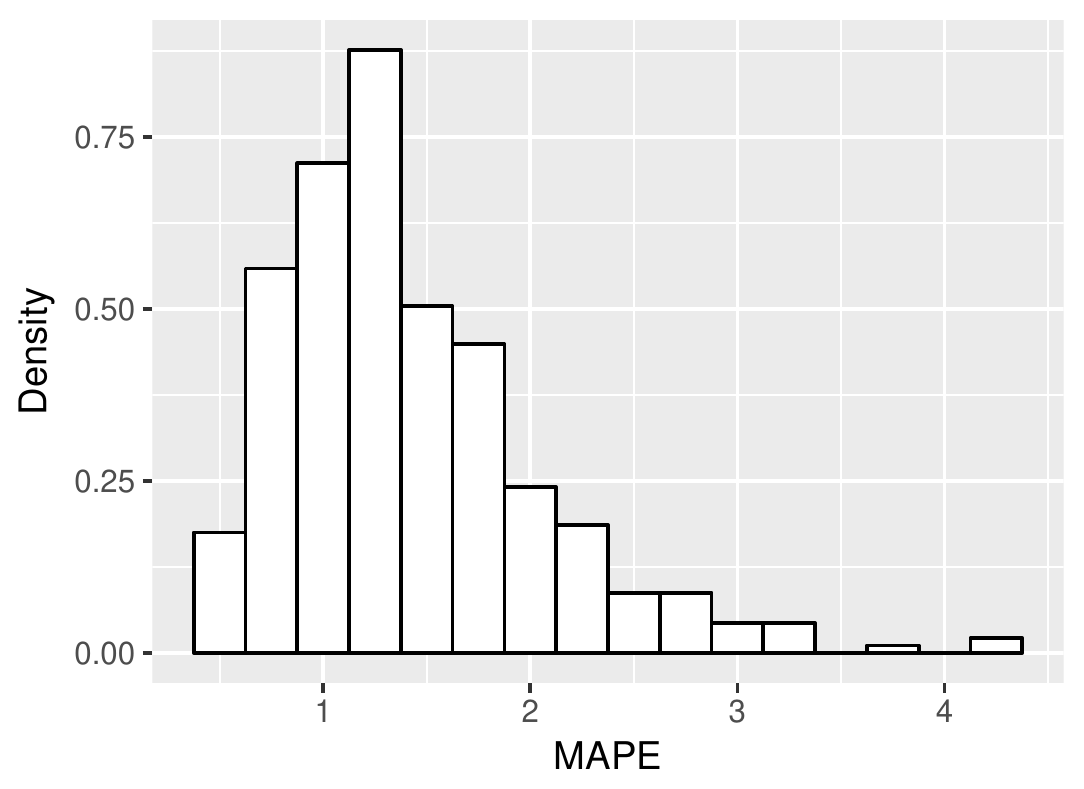}
\caption{Boxplot and histogram of MAPEs obtained by predicting half day of Spanish electricity demand during 2017.}
\label{fig:MAPEelectricity}
\end{center}
\end{figure}

Our forecasting exercise consisted in predicting half day of the 356 days of 2017.
Boxplot and histogram of mean absolute percentage errors (MAPE) obtained are shown is Figure \ref{fig:MAPEelectricity}.
Although they are different exercise, we remark that these MAPEs are smaller than those reported in the literature of functional methods for electricity demand forecasting.
In any case, our errors are very small, concentrated below $2\%$ and roughly they do not exceed $4\%$.

For testing the depth-based prediction bands, we consider again the three practitioner described on Section 3.2.
The results obtained are shown in Table 2.
They are even better than the results obtained for the PC processes.
The prediction band of the three simulated practitioner are very narrow, remaining high coverages.
To illustrate this case study, we show the plot from which the practitioner should select a $k$ value by considering two days with a different patter of electricity demand.

\begin{table}[]
\centering
\label{my-label}
\begin{tabular}{lllll|llll}
\cline{2-9}
                              & \multicolumn{4}{c|}{$\alpha = 0.05 $}                            & \multicolumn{4}{c}{$\alpha = 0.10 $}                       \\ \hline
\multicolumn{1}{l|}{Averages} & $C_{k^*}({\cal J})$ & $W_{k^*}({\cal J})$       & Prob. & $k^*$  & $C_{k^*}({\cal J})$ & $W_{k^*}({\cal J})$ & Prob. & $k^*$  \\
\multicolumn{1}{l|}{Pract. 1} & 0.921               & \multicolumn{1}{c}{0.215} & 0.937 & 17.711 & 0.897               & 0.197               & 0.890 & 14.940 \\
\multicolumn{1}{l|}{Pract. 2} & 0.921               & 0.215                     & 0.937 & 17.711 & 0.877               & 0.185               & 0.911 & 13.134 \\
\multicolumn{1}{l|}{Pract. 3} & 0.810               & 0.155                     & 0.958 & 9.019  & 0.711               & 0.125               & 0.921 & 6.000  \\ \hline
\end{tabular}
\caption{Results obtained by the same practitioner profiles described on Table 1 by predicting half day of Spanish electricity demand during 2017.}
\end{table}

%\begin{table}[]
%\centering
%\small
%\label{tab:electricity}
%\begin{tabular}{@{}lcc@{}}
%\toprule
%\toprule
%\multicolumn{1}{c}{\textbf{}} & {\textbf{Algorithm 1}}&{\textbf{Algorithm 2}}
%\\ \midrule
%\textbf{depth percentile on $[0, 720]$} & 0.977 & 0.980 \\
%\textbf{depth percentile on $(720, 1440]$} & 0.641 & 0.791 \\
%\bottomrule
%\end{tabular}
%\caption{ESTADÍSTICOS PARA EL "PLUG-IN" ESTIMATOR.}
%\label{tab:elec}
%\end{table}

\begin{figure}[]
\begin{center}
\includegraphics[height=6cm, width=16cm]{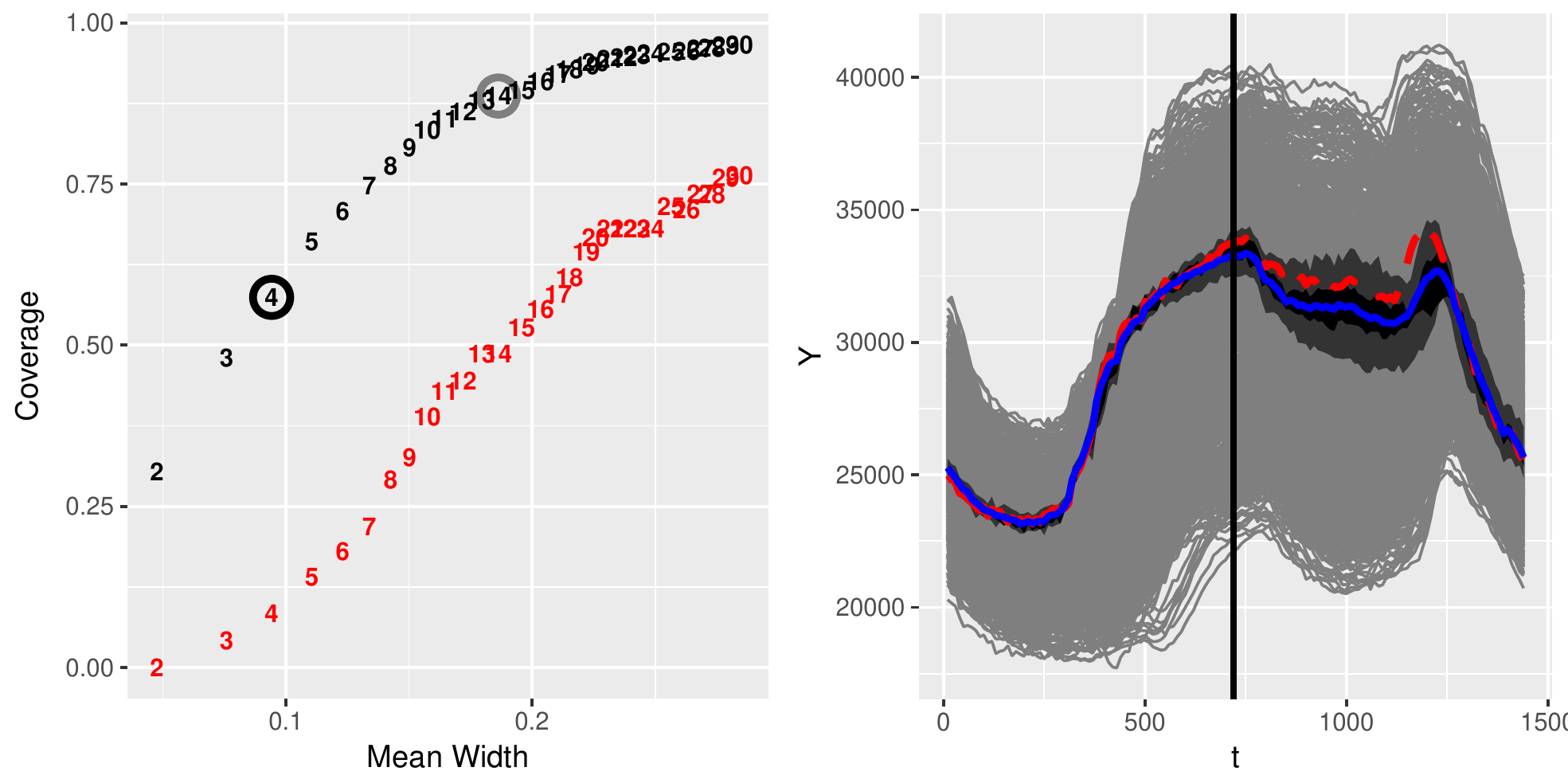}
\includegraphics[height=6cm, width=16cm]{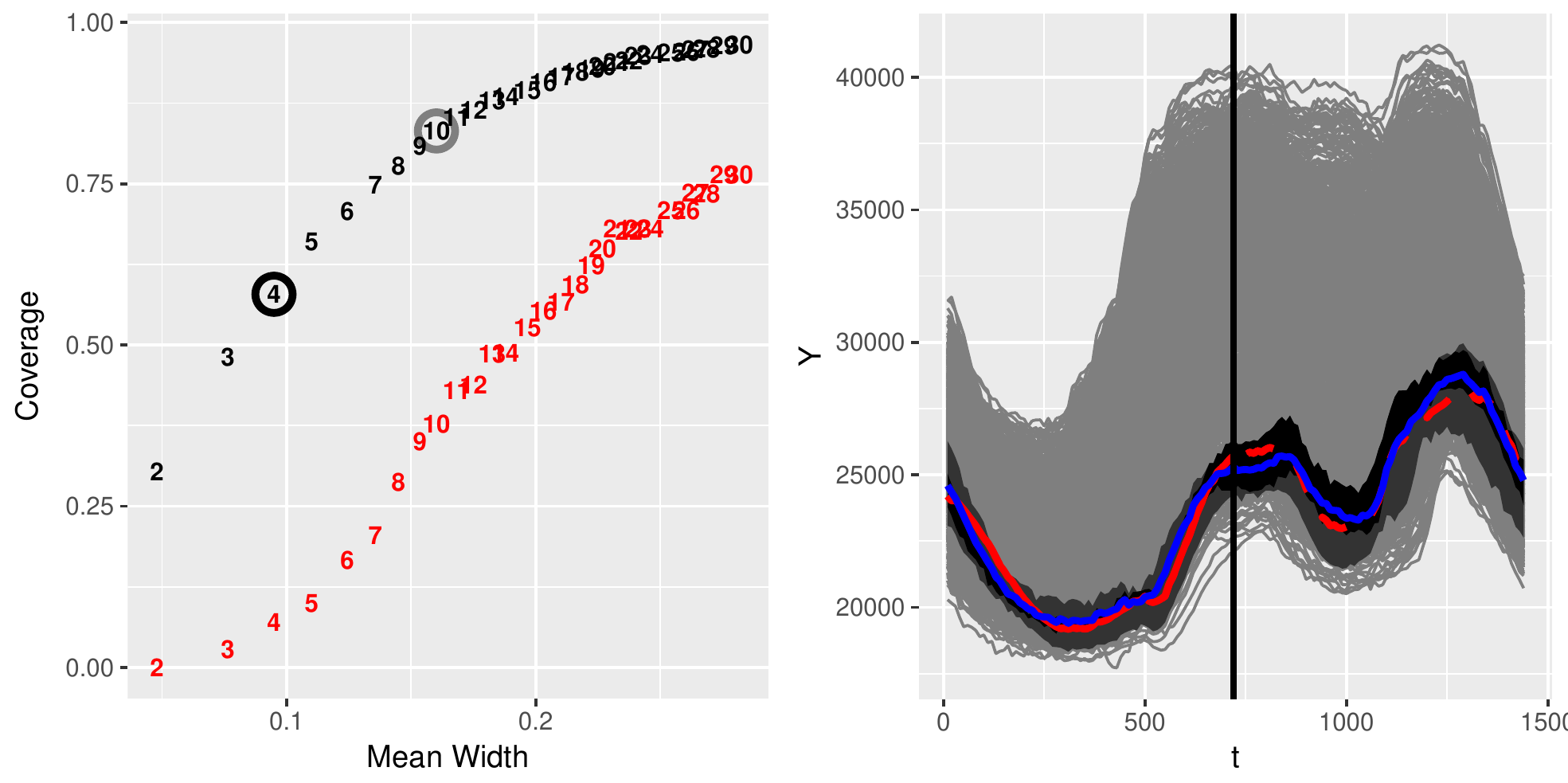}
\caption{Two examples of predicting exercises for daily Spanish electricity demand. Focal curve (in red) corresponds to: Top panel, a standard working day (Tuesday, Octuber 4th, 2017). Bottom panel, a Christmas day (December 25th, 2017).}
\label{fig:PredictionBandExample}
\end{center}
\end{figure}

As illustration, we show in top panel of Figure~\ref{fig:PredictionBandExample} the band corresponding to Tuesday, November 29th, 2016 by using Algorithm~\ref{alg:Alg1} and $k=10$.
Notably, all the curves used to envelope the focal come from months between November and February, however they correspond to different working days and years. 

\section{Conclusions}
\label{sec:conclusion}

Besides of providing point and band accurate forecast, the approach introduced here offers a new insight for functional time series forecasting.
In contrast with other methods, our method is completely empirical driven without attending to any statistical model that could explain the mechanism of data generation. 

The depth-based approach works with periodically correlated processes, that seem to give a wide pallet of test datasets for the functional time series analysis, and with case studies.
Beyond prediction, the tools we have discussed reveal qualitative aspects of the data structure of electricity demand.
For example,  focal-curve curve envelopes link weekends with holidays as well as atypical days with  past outliers.
Although these relations may appear somewhat naive, others could be not obvious.
In general, envelopes can be useful for discovering common features among periods.
The approach also introduce a tool for tuning coverage versus width in band forecasting.
This allows to practitioners to compute tight predictions bands that may preserve the shape to the curve to predict by diminishing coverage and fixing confidence level.
An alternative to other approaches in which the width band may be reduced only by losing confidence.
%As far as we know, this is the only method where the practitioner has access to control simultaneously coverage, width band and confidence level.
The main limitation of our approach is it is designed to functional time series with periodic rhythm and without trend.
If the effect of trend is negligible during a single period, one can subtract the functional mean to apply the method but there is no way of correcting the absence of periodic structure.
Despite this, many real systems generate data which are mixtures of randomness and periodicity, making that the range of applications be wide.

To end, we remark about the computational efficiency of the method. 
The data sets we have considered in this work entail a big computational challenge; more than one thousand curves observed in more than one hundred points. 
These data quantities are not only hard to be processed quickly but even they suppose a problem of capability of being processed.
In fact, regression-based methods require operations that may be intractable for sample sizes as we have considered. 
The methodology presented here is based on the Modified Band Depth and, as \cite{FastBD} argue,  this allows to rank million curves in only tens of seconds
making our algorithm efficient even for large data sets than the considered in our experiments.

\bibliographystyle{apalike}
%\nocite{*}
\bibliography{ref}

\newpage
\noindent
\large{\bf{Appendix}}

\algdef{SE}[SUBALG]{Indent}{EndIndent}{}{\algorithmicend\ }%
\algtext*{Indent}
\algtext*{EndIndent}
\begin{algorithm}
\caption{Input: ${\cal Y}, {y}_{n+1}$/ Output: $\mathcal{J}$}
\label{alg:Alg1}
\begin{algorithmic}[]
\small
\State 
\State \textbf{Initialize} ${\cal J} = \emptyset$
\While{ size of ${\cal Y}\setminus{\cal J} \geq 2 $ }
	\State Let $y_{(k)}$ be the $k$th-nearest curve to $y_{n+1}$ from ${\cal Y}\setminus{\cal J}$.
	\State ${\cal N} = y_{(1)}$ and $m=0$
	\For{$k\geq 2 $ }
	\State ${\cal N}^k = {\cal N} \cup \{y_{(k)}\}$
	\State $\lambda_k = \lambda \left( \{ t \in [0,q] : \min_{y \in {\cal N}^k} y(t) \leq y_{n+1}(t) \leq \max_{y \in {\cal N}^k} y(t) \} \right)$
	\If{$ \lambda_k> m$}
	\State ${\cal N} ={\cal N}^k$ and $m = \lambda_k$
	\EndIf
	\EndFor
		\State Let $p_0$ be the percentile of $D_{[0,q]}\left(y_{n+1}, {\cal J}^+\right)$ in $\{D_{[0,q]}(y, {\cal J}^+): y \in {\cal J}^+ \}$
		\State Let $p_1$ be the percentile of $D_{[0,q]}\left( y_{n+1}, {\cal J}^+ \cup {\cal N} \right)$ in $\{D_{[0,q]}(y, {\cal J}^+\cup {\cal N}): y \in {\cal J}^+\cup {\cal N} \}$
	    \If{ $p_1 \geq p_0$}
		\State ${\cal J} = {\cal J} \cup {\cal N}$ 
		\Else
		\State ${\cal Y} = {\cal Y} \setminus {\cal N}$
	   \EndIf
\EndWhile
\end{algorithmic}
\end{algorithm}

\end{document}